\documentclass[
    ,final            
  ]
  {aipproc}

\layoutstyle{6x9}

\newcommand{\rpv}{\mbox{$\not \hspace{-0.10cm} R_p$}}
\newcommand{\MET}{\mbox{$\not \hspace{-0.11cm} E_T$}}
\newcommand{\lle}{\mbox{$LL\bar{E}$}}
\newcommand{\lqd}{\mbox{$LQ\bar{D}$}}

\newcommand{\chiO}{\mbox{$\tilde{\chi}^0_1$}}

\usepackage{amssymb}

\begin{document}
\hfill\mbox{FERMILAB-CONF-06-367-E}
\\[-5ex]
\title{Search for $R$-Parity violating Supersymmetry with the D\O\ Detector}


\author{Christian Autermann \\for the D\O\ collaboration}{
  address={III. Phys. Inst. A, RWTH Aachen, Germany, auterman@fnal.gov}
}

\begin{abstract}
  Searches for R-parity violating supersymmetry with the D\O\ detector at the
  Fermilab Tevatron $p\bar{p}$-collider are presented. In the case of non-zero
  \lle\ couplings $\lambda_{ijk}\gtrsim 0.01$, multi-lepton final states,
  and for a small coupling $\lambda_{122}\ll 0.01$ di-muon final states are studied.
  The case of non-zero \lqd\ coupling $\lambda'_{211}$ leads to final
  states with two muons and jets. 
  
  A total integrated luminosity of $0.38$~fb$^{-1}$ collected between April
  2002 and August 2004 is utilized. The observed numbers of events are in
  agreement with the Standard Model expectation, and limits on \rpv\
  supersymmetry are derived, extending significantly previous bounds.
\end{abstract}

\maketitle


\section{Introduction}

Supersymmetry (SUSY) predicts the existence of a new particle for every
standard model (SM) particle, differing by half a unit in spin.  The quantum
number $R$-parity~\cite{farrar}, defined as $R = (-1)^{3B+L+2S}$, where $B$,
$L$ and $S$ are the baryon, lepton and spin quantum numbers, is $+1$ for SM
and $-1$ for SUSY particles. Often $R$-parity is assumed to be conserved,
which leaves the lightest supersymmetric particle (LSP) stable. However, SUSY
does not require $R$-parity conservation.
If $R$-parity violation (\rpv) is allowed, the following trilinear and
bilinear terms appear in the superpotential~\cite{rpv}:
\begin{eqnarray}
W_{\sl\rpv}&=& \frac{1}{2} \lambda _{ijk} L_i^{\alpha} L_j^{\beta} \bar{E}_k
            + \lambda ' _{ijk} L_i^{\alpha} Q_j^{\beta} \bar{D}_k 
            + \frac{1}{2} \lambda '' _{ijk} \bar{U}_i^{\xi} \bar{D}_j^{\psi} \bar{D}_k^{\zeta}
            + \mu_i L_i H_1
\label{super}
\end{eqnarray}
where $L$ and $Q$ are the lepton and quark SU(2) doublet superfields and
$\bar{E}$, $\bar{U}$, $\bar{D}$ denote the singlet fields. The indices have
the following meaning:  $i,j,k = 1,2,3$ = family index; $\alpha, \beta = 1,2 $
= weak isospin  index; $\xi, \psi, \zeta = 1,2,3 $ = color index.  The
coupling strengths are given by the Yukawa coupling constants $\lambda,
\lambda'$ and $\lambda''$. The last term, $\mu_i L_i H_1$, mixes the lepton
and the Higgs superfields.  The $\lambda$ and $\lambda'$ couplings give rise
to final states with multiple leptons, which provide excellent signatures at
the Tevatron. All analyses presented here require that only one \rpv\ coupling
is of significant size, since strict limits exist on the product of two
couplings, i.e. from the proton lifetime~\cite{referbound}. A detailed review of \rpv\ SUSY is given in \cite{barbier}.

The data for this analysis were recorded by the D\O\ detector between April 2002
and August 2004 at a center-of-mass energy of $\sqrt{s}=1.96$~TeV. The
integrated luminosity corresponds to $380\pm 25$~pb$^{-1}$. A detailed
description of the D\O\ detector can be found in \cite{d0det}. 


\subsection{Gaugino pair and associated production}

\begin{figure}[b]
  \begin{tabular}[]{cc}
  \includegraphics[height=.3\textheight]{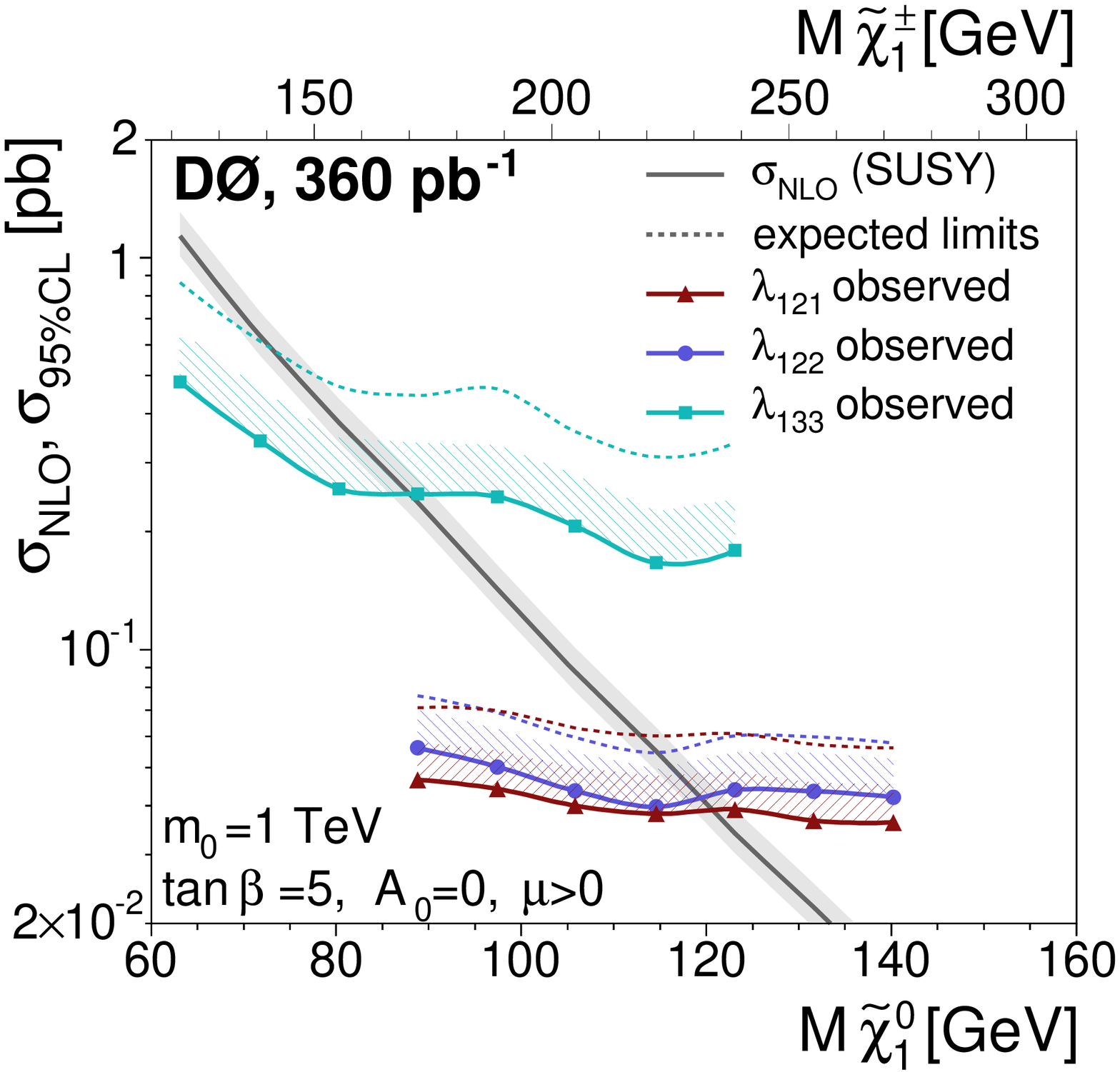} &
  \includegraphics[height=.3\textheight]{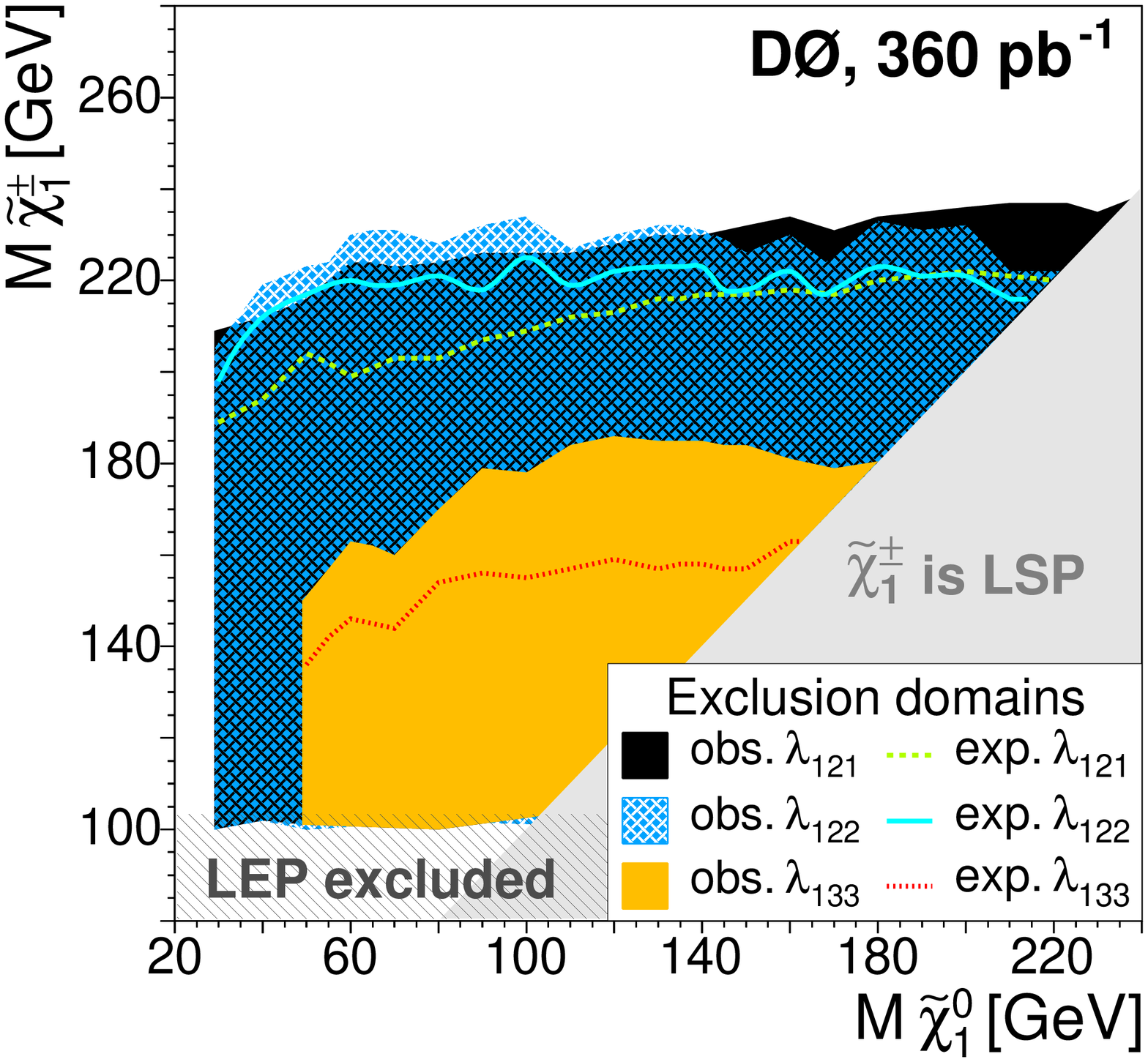}
  \end{tabular}
  \caption{Cross section limit for three different \lle\ couplings compared to
	   the mSUGRA cross section prediction (left). Within mSUGRA with
	   relaxed gaugino GUT relation  exclusion contours for each
	   coupling $\lambda_{121}$, $\lambda_{122}$, $\lambda_{133}$  in the
	   m($\tilde{\chi}^0_1$)--m($\tilde{\chi}^\pm_1$) plane are derived
	   (right).~\cite{Abazov:2006nw}}
  \label{pair_long}	   
\end{figure}

The charginos and neutralinos are produced in pairs or
associated. The produced sparticles (cascade) decay to the lightest
neutralino \chiO. When assuming a non-zero \lle\ coupling, then this
neutralino decays into two charged leptons and one neutrino by violating
R-parity. The final state therefore contains at least four charged leptons
and two neutrinos which lead to missing transverse energy (\MET) in the detector.

This analysis requires the prompt decay of the lightest neutralino
$\tilde{\chi}^0_1$, so that all particles originate from the same vertex,
leading to the constraint that the corresponding \lle-coupling ($\lambda_{121}$,
$\lambda_{122}$, or $\lambda_{133}$) is larger than $\gtrsim 0.01$.  For best
acceptance, only three charged leptons are required to be identified. Three
different analyses are performed
depending on the flavors of the leptons in the final state $eel$, $\mu\mu l$,
$ee\tau$, with $l=e,\mu$. All three analyses
are optimized separately using SM and signal MC simulations.
Details of the selection and the analyses can be found in \cite{Abazov:2006nw}.

Since no evidence for \rpv-SUSY is observed in tri-lepton events, the analyses
are combined. Upper limits on the chargino and neutralino pair production cross
section are set. Lower bounds on the masses of the lightest neutralino and the
lightest chargino are derived in mSUGRA and in an MSSM scenario with heavy
sfermions, but assuming no GUT relation between the gaugino masses $M_1$ and
$M_2$. All limits as shown in Fig.~\ref{pair_long} are the most restrictive to
date.

\pagebreak

\subsection{Neutral long lived particles (NLLP)}
Here, a {\it small} \lle\ coupling $\lambda_{122}$ is assumed leading to
long neutralino $\tilde{\chi}^0_1$ lifetimes and a displaced di-muon vertex. 
The primary vertex was reconstructed using all tracks, except
those associated with muons. The primary vertex is required to be within
$0.3$~cm of the beamline in $x$ and $y$ and within $60$~cm of the detector
center in $z$. Two muons must originate from the same secondary vertex, that is displaced
$5-20$~cm with respect to the primary vertex. Details can be found
in~\cite{Abazov:2006as}.

\begin{figure}[hb]
  \includegraphics[height=.3\textheight]{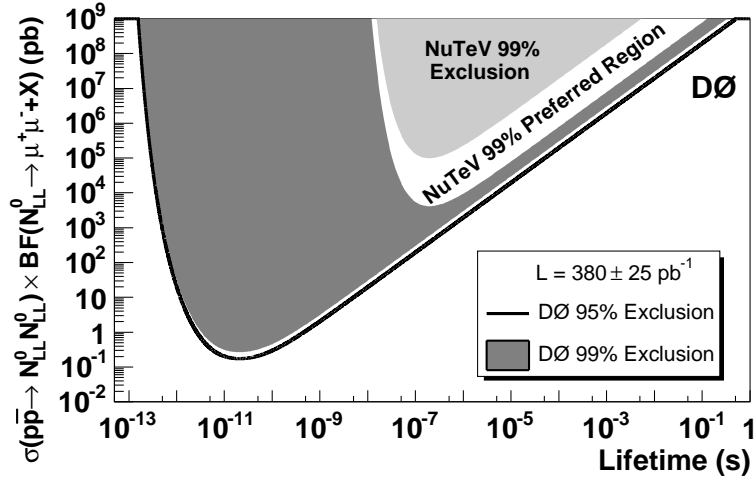} 
  \caption{Cross section limit with 95\% CL. The interpretation of the NuTeV
  excess in the di-muon channel arising from the decay of neutral long lived particles is shown as
  well and can be excluded.~\cite{Abazov:2006as}}
  \label{pair_short}
\end{figure}

This analysis is sensitive to neutral, long-lived particles decaying to $\mu\mu$+X.
The background is estimated to be $0.75 \pm 1.1$ (stat) $\pm 1.1$ (syst) events.
No events were selected and a limit on the product of NLLP pair production cross section times
decay branching fraction into $\mu\mu+X$ is set as a function of the 
lifetime.
The 95\% CL cross section limit for a mass of $10$
GeV and a lifetime of $4 \times 10^{-11}$~s is $0.14$~pb. The result as shown in
Fig.~\ref{pair_short} excludes an interpretation of the NuTeV excess~\cite{nutev} of di-muon
events in a large class of models.

\subsection{Resonant Second Generation Slepton Production}
The \lqd\ coupling offers the opportunity to produce sleptons 
in $p\bar{p}$-collisions as resonances. For a non-zero coupling
$\lambda'_{211}$ this is either a smuon or a muon sneutrino. The
slepton cascade decays into the lightest neutralino \chiO\ and associated
leptons. The neutralino decays via the same $R$-parity violating coupling
$\lambda'_{211}$ into a $2^{nd}$ generation lepton and two jets. The cross
section is proportional to $(\lambda'_{211})^2$, so that limits on this
coupling can be derived.

Three resonant slepton channels (i)~$\tilde{\mu}\to\tilde{\chi}^0_1\,\mu$,
(ii)~$\tilde{\mu}\to\tilde{\chi}^0_{2,3,4}\,\mu$, and
(iii)~$\tilde{\nu}_\mu\to\tilde{\chi}^\pm_{1,2}\,\mu$ resulting in di-muon and
multi-jet final states for neutralino decays $\tilde{\chi}^0_1\to\mu q \bar{q}'$
are analyzed separately. For the further discrimination of the signal and the
Standard Model background, the analyses make use of the possibility to
reconstruct the neutralino $m(\tilde{\chi}^0_1)=m(\mu, q, q)$ and the
slepton $m(\tilde{l})=m(\mu, \mu, q, q)$ masses. More details can be found
in~\cite{Abazov:2006ii}. 

In the absence of an excess in the data, cross section limits on resonant
slepton production were set. To be as model independent as possible,  limits
with respect to the slepton production cross section times branching fraction to
gaugino plus muon are calculated.  The results are interpreted within the mSUGRA
framework with $\tan\beta=5$, $\mu<0$, and $A_0=0$ and an exclusion contour
w.r.t. $\lambda'_{211}$ is derived.  All three channels were combined to form
one limit for $q\bar{q}\to\tilde{l}$.   Lower limits for the slepton mass of
$210$, $340$ and $363$~GeV independent of other masses are obtained for
$\lambda'_{211}$ values of $0.04$, $0.06$ and $0.10$, respectively, a
significant improvement compared to previous results.

\begin{figure}[ht]
  \begin{tabular}[]{cc}
  \\
  \includegraphics[height=.3\textheight]{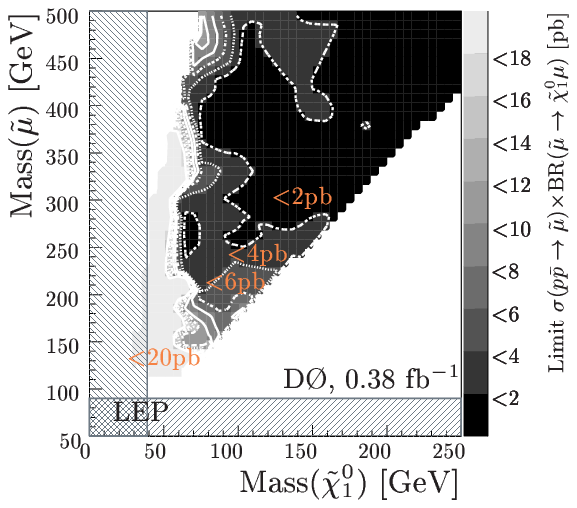} &
  \includegraphics[height=.3\textheight]{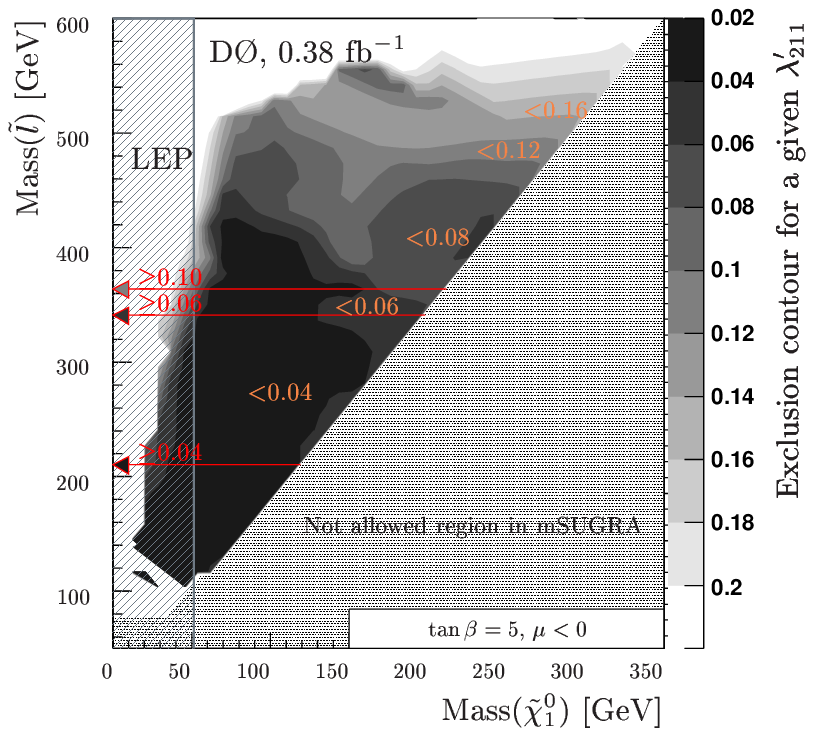}
  \end{tabular}
  \caption{Cross section times branching fraction limit for the process
	   $\tilde{\mu}\to\tilde{\chi}^0_1\mu$ (left). All three resonant slepton
	   production channels can be combined within mSUGRA and are
	   translated into an exclusion contour with respect to the \lqd\ coupling
	   strength $\lambda'_{211}$ (right).~\cite{Abazov:2006ii}}
\end{figure}


\end{document}